\documentclass[a4paper,fleqn,usenatbib]{mnras}
\usepackage[T1]{fontenc}
\usepackage{ae,aecompl}
\usepackage{newtxtext,newtxmath}
\usepackage{graphicx}
\usepackage{amsfonts}
\usepackage{amssymb}
\usepackage{colors}
\usepackage[normalem]{ulem}

\title{Implications for the missing low-mass galaxies (satellites) problem from cosmic shear}
\author[Jimenez R., Verde L., Kitching T. D.]{Raul Jimenez$^{1,2}$\thanks{raul.jimenez@icc.ub.edu}, Licia Verde$^{1,2}$, Thomas D. Kitching$^3$\\
$^1$ICC, University of Barcelona, Marti i Franques 1, 08028, Barcelona, Spain.\\
$^2$ICREA, Pg. Lluis Companys 23, 08010 Barcelona, Spain.\\
$^3$Mullard Space Science Laboratory, University College London, Holmbury St Mary, Dorking, Surrey RH5 6NT, U.K.\\
}

\newcommand{\be}{\begin{equation}}  \newcommand{\ee}{\end{equation}}
  \newcommand{\ba}{\begin{eqnarray}}
\newcommand{\ea}{\end{eqnarray}}

\def\gs{\mathrel{\raise1.16pt\hbox{$>$}\kern-7.0pt %
\lower3.06pt\hbox{{$\scriptstyle \sim$}}}}         %
\def\ls{\mathrel{\raise1.16pt\hbox{$<$}\kern-7.0pt %
\lower3.06pt\hbox{{$\scriptstyle \sim$}}}}         %

\begin{document}

\voffset=-0.25in 

\maketitle

\begin{abstract}

The number of observed dwarf galaxies, with dark matter mass $\lesssim 10^{11}$ M$_{\odot}$ in the Milky Way or the Andromeda galaxy does 
not agree with predictions from the successful $\Lambda$CDM paradigm. To alleviate this problem 
 a  suppression of 
dark matter clustering power on very small scales has been conjectured. However, the abundance of 
dark matter halos outside our immediate neighbourhood (the Local Group) 
 seem to agree with the  $\Lambda$CDM--expected abundance. 
Here we  connect  these problems to  observations of weak lensing cosmic shear, 
pointing out that cosmic shear can make significant statements about the missing satellites problem in a statistical way. 
As an example and pedagogical application we use
recent constraints on  small-scales power suppression from  measurements of the CFHTLenS data.
 We find that, on average, in a  region of $\sim $Gpc$^3$ there 
is no significant small-scale power suppression. This  implies
that suppression of small-scale power is not a viable solution to the `missing satellites problem' or, 
alternatively, that on average in this volume there is no `missing satellites problem' for dark matter masses $\gtrsim 5 \times 10^9$ M$_{\odot}$. 
Further analysis of current and future weak lensing surveys will 
probe  much smaller  scales, $k > 10h$ Mpc$^{-1}$ corresponding roughly to masses $M < 10^9 M_{\odot}$.

\end{abstract}


\section{Introduction}
\label{Introduction}
The observed abundance of satellite galaxies around the Milky Way with dark matter masses less than the 
Large Magellanic Cloud ($8 \times 10^{10}$ M$_{\odot}$) does not agree with the number of 
corresponding dark matter halos predicted by detailed N-body simulations of the current $\Lambda$CDM paradigm (\citet{Klypin,Moore}; 
see also some more recent high-resolution simulations in \citet{Diemand,acquarius}). This is because in cosmological simulations 
that incorporate only gravity and collisionless cold dark matter, simulated halos retain large amounts of substructure 
formed by earlier, smaller-scale, collapse, predicting hundreds of sub-halos in contrast to the $\sim 10$ observed satellites 
of the Milky Way. This has been referred to as the `missing satellites problem'. Several solutions have been proposed to 
reconcile this problem with the $\Lambda$CDM predictions ranging from the original proposal by \cite{Heavens} of the 
existence of low-mass dark galaxies, to the suppression of power at small-scales either 
in the initial conditions e.g.,~\citet{kamionkowskiLiddle00,ZentnerBullock03}, or by changing the properties of the 
dark matter -- to be interacting, or not cold -- e.g.,~\cite{Colinetal00, Bodeetal01,Strigarietal07}, to 
baryonic processes, or processes proposed to render these satellites dark (i.e. have a very low stellar to dark matter mass ratio; 
see e.g the review by \cite{Bullock10} and also \cite{Verde02}).

It is important to note that the missing satellite problem is mostly confined to observations of the Local group. Although 
some indications hint towards dwarf galaxies not being missing in systems beyond the 
local group \citep{Cote,Read,Fontana17}, and image flux ratios in images of strongly lensed galaxies indicate the presence 
of substructure  (e.g., \cite{Dalal:2001fq,Dobler:2005pk,Vegetti:2012mc,H1,Vegetti:2014lqa,H2,H3,Diaz}) in broad 
agreement with $\Lambda$CDM predictions, we have little direct information  about the amount of 
substructure (satellites) and their mass distribution in other galaxies. A closely related problem is the 
fact that the observed faint end of the luminosity function has a shallower slope \citep{Blanton01,Panter} than 
predicted from high-resolution $\Lambda$CDM simulations; however this relies on assuming a constant 
mass-to-light ratio (e.g.,~\cite{Jenkins01}). Stellar feedback is again often used in semi-analytic models to 
suppress star formation in halos with shallow potential wells e.g.,~\cite{HopkinsetalFIRE2013}. Finally, the determination 
of halo masses bygalaxy-galaxy lensing (e.g. \cite{Mandelbaumetal06} in SDSS), gives satellite fractions consistent 
with those of the $\Lambda$CDM model.

It should be clear that  explanations for the missing satellite problem can be broadly divided in two classes: {\it i)}  those that suppress the (linear) matter power spectrum on small scales as  to suppress the number of small halos compared to the predictions of  power law power spectrum ($\Lambda$CDM) model and {\it ii)} those that leave the $\Lambda$CDM power spectrum untouched, but simply hide these halos by leaving  them  dark. Therefore, indirect statistical information could be gathered by measuring the matter power spectrum on small scales over a representative volume of the Universe; 
 this quantity  directly determines the abundance (and mass function) of satellites under the assumption that 
gravity (and not baryonic physics) is the dominant process at play. If the power spectrum is consistent with a power law $\Lambda$CDM then the remaining plausible  explanation for the missing satellite problem is either that for environmental reasons they   are only missing in our galaxy but not elsewhere, or that they are indeed ubiquitous but  dark.
The Lyman$-\alpha$ forest  has been used to probe the  neutral hydrogen density and infer the dark matter power spectrum, 
thus probing scales that today are highly non-linear, see e.g.,  \cite{Viel:2007mv, Boyarsky:2008xj}. Until now, no study has 
been able to recover  directly the low redshift matter power on sufficiently small scales ($k\sim 5h$Mpc$^{-1}$) for a 
representative volume, thus providing a sufficiently fair sample of the small scale power in the late-time Universe. Here 
we point out that cosmic shear measurements can provide such a data set. 

As an example analysis, that serves as an indicator of what could be done with larger and better data sets, 
we will use a recent analysis that recovers the power spectrum of matter from a spherical-Bessel analysis of weak lensing data 
in CFHTLenS. We find that this example analysis provides support for it being consistent with that of $\Lambda$CDM and not showing 
any deficit of small-scale power. This in turn can be used to constrain cosmological models or dark matter properties that 
suppress small scale power and, if gravity is the dominant force at play, to quantify the predicted abundance of low mass halos.

\section{Method and data}
\label{Method}
Here we describe how cosmic shear measurements can be used to infer information on the low-mass galaxy population. 
We will use a particular implementation of cosmic shear \citep{KitchingI} as an example of the type of analysis that can result in 
statements on the low-mass galaxy population.  We use this because it is a type of analysis that can relatively 
cleanly identify what scales (i.e., wavenumbers $k$) contribute to the observed signal in the  cosmic shear statistic.

Weak lensing of galaxy images, the effect where the 
observed shape of galaxies is distorted by the presence of mass 
perturbations along the line of sight, is a particularly interesting 
probe of matter distribution in the Universe. This is because 
the distortion - a change in the third eccentricity, or third flattening (known as `ellipticity'), 
and size of galaxy images - depends on perturbations in the total matter density which, 
because we live in an apparently dark matter-dominated Universe,  
is in principle is sensitive to the dark matter power spectrum directly. 
Accessing the matter power spectrum through weak lensing measurements 
results in a statistic that contains a wealth of cosmological information.
Here we use the recently measured spherical-Bessel power spectrum of the weak lensing effect, a
statistic known as `3D cosmic shear', and use this to explore differences between the 
inferred matter power spectrum and that predicted within a standard, power law cold dark matter-only $\Lambda$CDM model.

The small-scale power spectrum is very poorly understood at the current time for two reasons. The 
first is that highly non-linear dark matter clustering is not well modelled. Analytic approaches based on perturbation theory are only valid on mildly non-linear scales $k \sim {\cal O}(0.1)$ at $z=0$,  numerical simulations and phenomenological fitting formulae must be employed. Current simulations  and fitting emulators
are  precise to a few percent up to scales of 
 $k=5$ $h$Mpc$^{-1}$ (e.g., \cite{Lawrence10, Lawrence17}), phenomenological fitting formulae are accurate to 10\% down to $k=10$ $h$ Mpc$^{-1}$ e.g., \cite{Takahashi}. The second is that 
the $\Lambda$CDM paradigm could break down at small scales and new physical processes 
could be present, for example modified gravity models, neutrino physics, and warm dark matter 
models interaction in the dark sector all have potentially detectable signatures at scales smaller than $1$ Mpc. Thus measuring the matter power spectrum at  small  scales would yield key  information about all these processes. This is in principle accessible by  present and  forthcoming analyses of weak lensing surveys.

In a previous paper \citep{KitchingI}, we described the method to infer the 
power spectrum from current weak lensing data. The data used was CFHTLenS, \citep{Erben13,Heymans13}, 
which is a $154$ square degree optical 
survey (over four fields W1, W2, W3, W4) in $griz$ bands, with weak lensing shape measurements \citep{Miller} and photometric 
redshift posterior probabilities \citep{Hilde}. There we presented a measurement of the power spectrum for 
matter in the range $0.001 \leq k \leq 5 h$Mpc$^{-1}$ (Fig. 6). It should be noted that in  \cite{KitchingI} we 
adopted the best fit marginal values from the \cite{Planck} cosmology for the cosmological parameters that are not 
explicitly varied. This is because the CFHTLenS data do not have enough statistical power to leave all 
cosmological parameters free and constrain them independently of the CMB data. To be more specific, the model parameters 
that we fit to the data are divided in three parts: the cosmological model, the baryonic feedback model, and the 
parameters for photometric redshift systematic effects. In \cite{KitchingI} we explore  the feedback and 
photometric redshift parameters  within  physical constraints. On the other hand, we adopt the  
Planck best fitting values for  $\Omega_B$ the dimensionless density of baryons, $H_0$ the Hubble constant and 
the spectral index of of the initial density perturbations $n_s$.  We  assumed that the dark energy is a 
cosmological constant and a flat geometry such that $\Omega_{\Lambda} = 1 - \Omega_M$. Finally, we assumed the total 
sum of neutrino masses to be zero. 
Although a power law primordial power spectrum is 
assumed with a given spectral slope, the freedom allowed by the parameterisation of the baryonic 
feedback model (see \cite{KitchingI} sec. 2.4.2) leaves abundant freedom to the reconstructed shape of the power spectrum. Effectively this yields a  minimally parametric reconstruction of  the  shape of the matter power spectrum at small scales.
\begin{figure}
\centering
    \includegraphics[angle=0,clip=,width=\columnwidth]{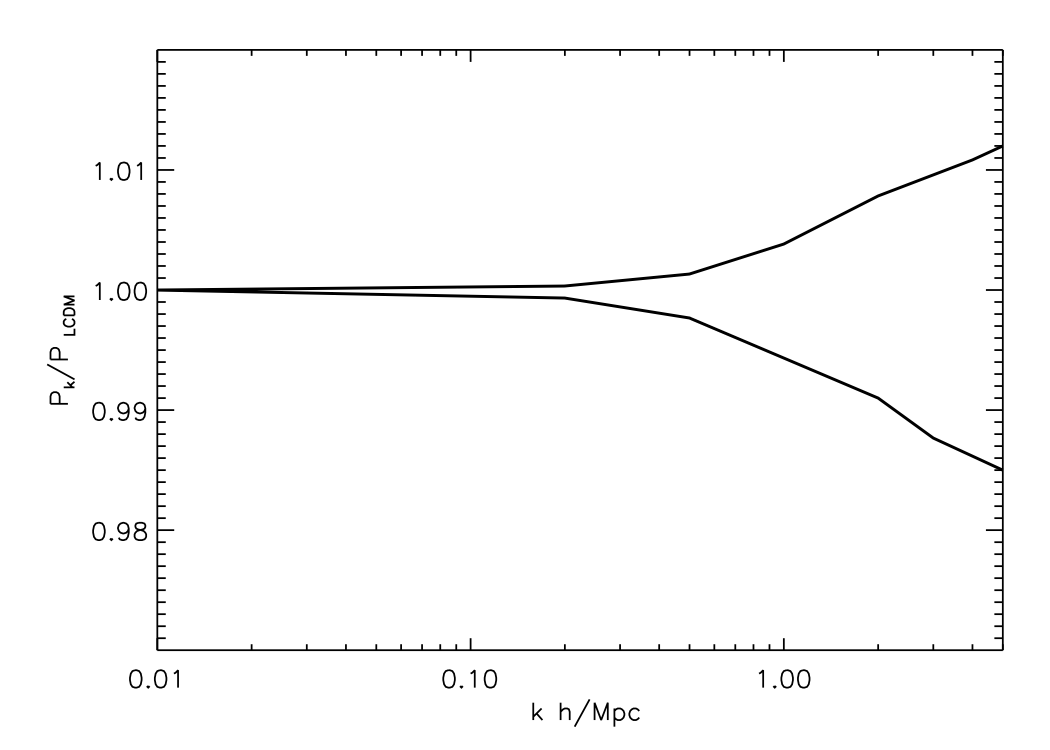}
\caption{Ratio of the power spectrum of matter from CFHTLenS as a function of scale and the Planck collaboration best fit $\Lambda$CDM model;
the two lines show the 68\% confidence regions for the CFHTLenS power spectrum. Note that both agree at
better than few \% accuracy. Because we only recovered the power spectrum up to scales of $k= 5 h$ Mpc$^{-1}$ we do not know how it looks for 
larger values of $k$; because of this we have simply assumed that it has a sharp cut-off 
as this will be the worst-case scenario when investigating the abundance of low mass dwarf galaxies.}
 \label{fig1}
\end{figure}

\begin{figure}
\centering
    \includegraphics[angle=0,clip=,width=\columnwidth]{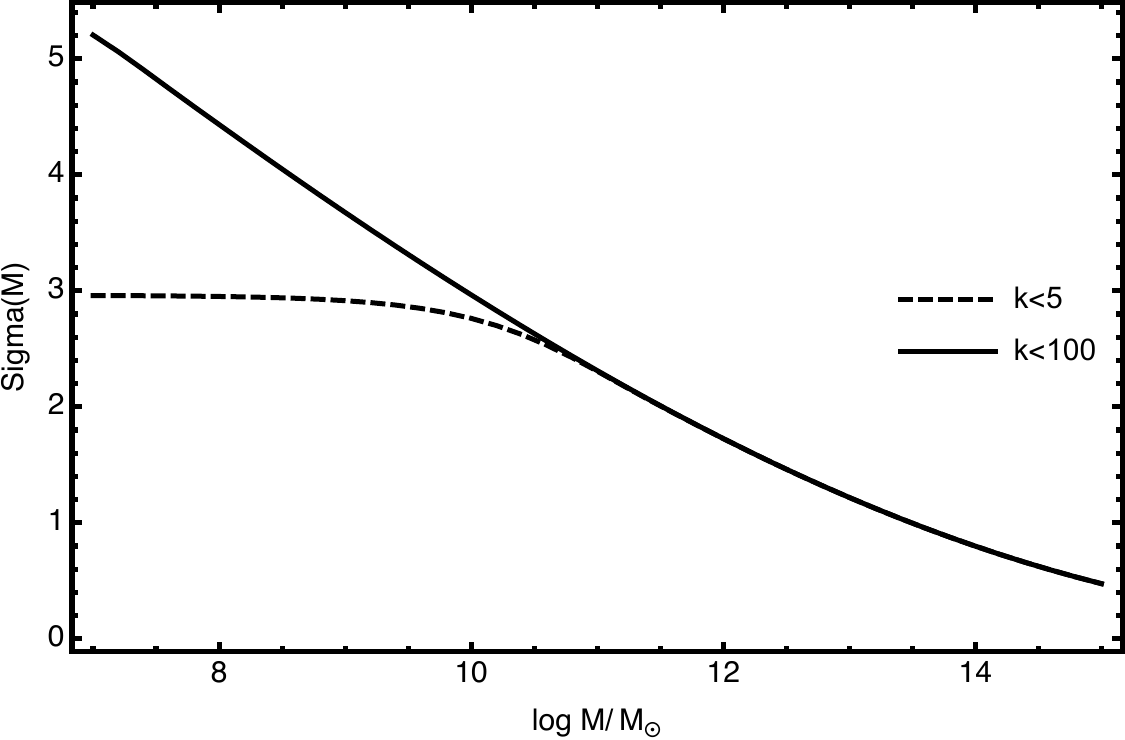}
 \caption{The rms mass fluctuation as a function of the enclosed mean mass $M$ for two power spectra, one truncated at $k>5\,h$ Mpc$^{-1}$ and one at  $k>100\,h$ Mpc$^{-1}$. As expected, 
for the CFHTLenS case adopted here, for masses lower than $5 \times 10^9$ M$_{\odot}$, the rms fluctuations are lower and thus will be the abundance  of these objects.}
 \label{fig2}
\end{figure}

\begin{figure}
\centering
    \includegraphics[angle=0,clip=,width=\columnwidth]{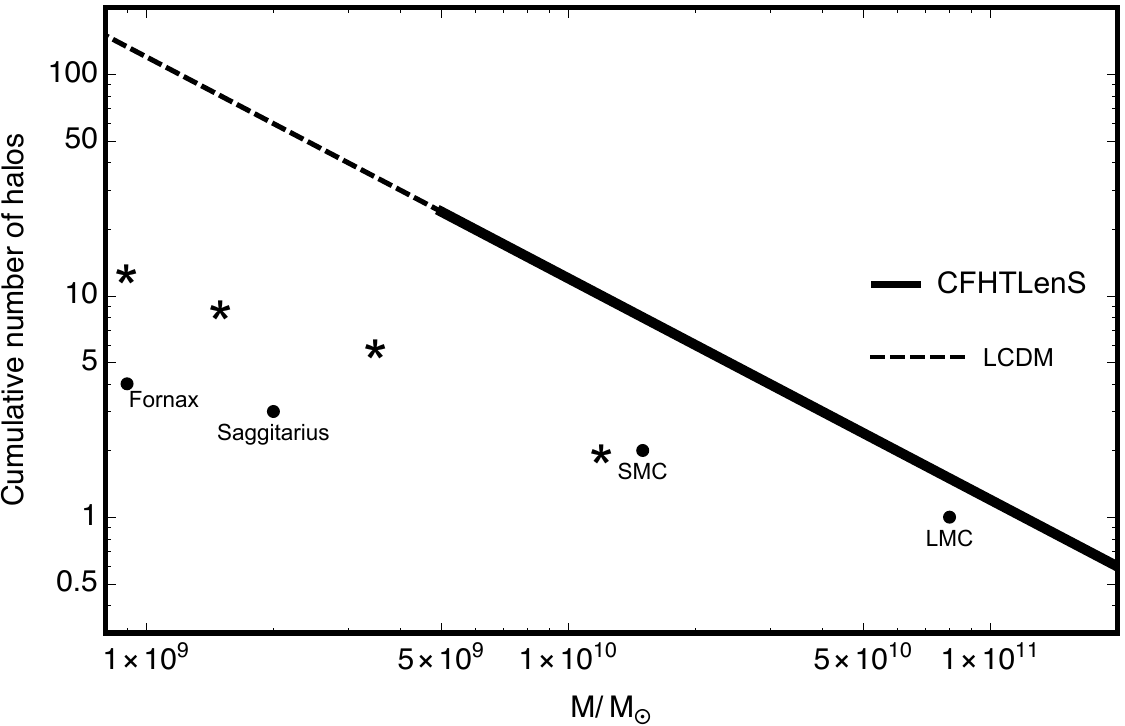}
 \caption{The cumulative number of halos as a function of their mass for the Milky Way (dots) and M31 (stars) and that predicted from the 
CFHTLenS power spectrum reconstruction (solid line). Note that unlike for the Milky Way and M31, in a statistical sense as reconstructed from lensing data, 
there is no missing number of satellites in the volume of $\sim Gpc^3$  probed by the CFHTLenS data compared to the $\Lambda$CDM prediction (dashed line). The error on the reconstructed case 
is of order the size of the thickness of the line.}
 \label{fig3}
\end{figure}
The question we want to address is the following: does the   the reconstructed  power spectrum  allow,  within its uncertainty, a  small scales power suppression large enough to reduce significantly the predicted number  of satellites  and thus solve the missing satellites problem?


From the CFHTLenS 3D lensing  estimate of the power spectrum we compute
the predicted abundance of halos by following the standard spherical collapse approach, 
calibrated on the latest numerical simulations by \cite{coyote}.  In brief: the first step is to compute the linear-theory rms fractional mass 
fluctuation $\sigma(M)$ in spheres of radii\footnote{This is the initial Largangian radius, $R$, that encloses a mass $M=4/3\pi R^3\Omega_m\rho_c$ where $\rho_c$ denotes the Universe critical density. This should be computed at the redshift of interest which we take it to be $z=0$.}  that enclose a mass $M$ using a top-hat window function. 
This requires a linear  matter power spectrum, while observations yield the non-linear one.  We convert
between non-linear (which is what we obtained from the cosmic
shear-derived power spectrum) and linear matter power spectrum using the package HaloFit \citep{halofit,Takahashi}. This requires an
inversion process in HaloFit. It should be noted that the non-linear measured power spectrum is in good agreement with that predicted from a power law $\Lambda$CDM model.  Any deviations will be small. Therefore we  generate multiple
(1000) non-linear power spectra for a flat $\Lambda$CDM-like cosmologies
where we let the cosmological parameters free, including power spectrum spectra index and running
of the power spectrum, as to reproduce a possible suppression of power at small scales.  
The cosmological parameters we allow to vary are: the matter cosmological densities;
the Hubble constant and the running of the power spectrum, i.e. a
scale dependent index for the power law. The priors we adopt are
$5-\sigma$ of  the Planck errors for the $\Lambda$CDM best fitting model for the
cosmological densities; for $H_0$ we adopt a very wide 50 -- 100 km
${\rm s}^{-1} {\rm Mpc}^{-1}$ prior. Finally, for the running of the power spectrum,
we use a prior the range $ -0.05 < \frac{d n_s}{d ln k} < 0.05$. From this set of $1000$ non-linear power spectra we find the one best matching the observed one.

For the second step, which follows the same philosophy of the paper by \citet{kamionkowskiLiddle00}, in order to compute the abundance of objects above a certain mass we use the expression for the 
mass function provided by \citet{coyote}, for standard values of their collapse parameters and for 
our observed matter power spectrum from CFHTLenS. This cumulative number of halos is computed as
\begin{equation}
F( > M_{\rm low}) = {\rm erfc} \left [ \frac{\delta_c z_f}{\sqrt{ 2 [ \sigma^2(M_{\rm low})-\sigma^2(M_{\rm high})]}} \right ]
\label{eq:halos}
\end{equation}
where $M_{\rm low}$ is the mass in subhalos that at a formation redshift $z_f$ yields a mass $M_{\rm high}$, 
for $\delta_c$ we adopt a value of $1.7$.  

It is easy to  understand why, even by artificially setting the non-linear power to zero above $k=5\,h$ Mpc$^{-1}$,  the estimated mass function (which depends on  an integral of the linear power spectrum convolved with a low pass filter) does not deviates much from a $\Lambda$CDM one in the reported mass range ($M>5 \times 10^9$ M$_{\odot}$).  In the halo-Fit \citep{halofit} philosophy  non-linearities do not change the power, but map the wavenumbers  between linear and non linear ones, $k_{NL}$ and $k_{L}$ respectively, via  $k_{NL}=(1+\Delta^2(k_{NL}))^{1/3} k_{L}$, where $\Delta^2(k_{NL})$ is the dimensionless (non-linear) power.   Since at the median sources of the CFHTLens lenses \citep{KitchingCFHT}, $\Delta^2 \sim 50$  for $k_{NL}=5\,h$Mpc$^{-1}$,  $k_{L}$ is only a factor $\sim 2$ smaller. This means that we expect our prescription    to 
 start underestimating  the linear power at scales $k_{L}> 3\,h$Mpc $^{-1}$ and thus for masses somewhere below $10^{10}$ M$_{\odot}$. This helps understand why the effect on the cumulative mass function starts being evident at masses below $5 \times 10^9$ M$_{\odot}$, which is where we stop reporting the results (see Fig. 3).

\section{Results}
In Fig.~\ref{fig1} we show the allowed range of freedom in the small scale shape of the matter power spectrum obtained by the \citet{KitchingI} reconstruction.  The power spectrum on large  (linear) scales is by construction imposed to be that of the Planck marginal best fit. At  small scales the freedom allowed by the minimally parametric reconstruction results in an uncertainty band;
the two solid lines indicate the 68\% uncertainty range on the 
recovered power spectrum.  
Since we only recover the CFHTLenS spectrum up to $k= 5h$Mpc$^{-1}$, in what follows we assume that for larger $k$-modes there is a sharp suppression of 
power; this is to study the best possible case for matching the observed low abundance of small mass halos. 
Note that the uncertainty in our recovered power spectrum is $\sim 1$\% even when leaving abundant freedom to the small-scale power. 
The linear power spectrum is recovered through a repeated forward procedure involving halo fit as explained above. Note that there is no extrapolation to $k>5\,h$ Mpc$^{-1}$ since in the halo fit procedure the non-linear $k$ produced by the recipe is always larger than the linear one.
From this linear power spectrum we compute the predicted halo abundance as described in Section \ref{Method}. 
Fig.~\ref{fig2} shows $\sigma(M)$ as a function of mass for a power law $\Lambda$CDM power spectrum at all scales (for the best fit Planck parameters, as used in Fig.~\ref{fig1} and one cut at $k>5$Mpc/h (as used for the CFHTLenS case here). As expected, we do observe a lack of fluctuations in the presence of the artificial small scales power cutoff imposed, in this case the suppression is drastic for halos masses  below $5 \times 10^9$ M$_{\odot}$. No visible suppression can be seen above $10^{10}$ M$_{\odot}$-- where the  missing satellite problem still persists. Hence by artificially cutting power above $k=5\, h$Mpc$^{-1}$ we are being conservative.

In Fig.~\ref{fig3} we show the inferred cumulative abundance of objects above  mass $M$ as a solid thick line; this is our main result. 
For comparison, we also show the observed cumulative number of halos in the Milky Way as a function of their inferred dark matter mass; 
the latter has been derived from the observed circular speed as $v_c^3 = 10 M G H(z_f)$, $G$ is the Newton constant and $H$ the Hubble parameter at 
collapse of the dark matter halo, which we adopt to be $z_f =1$. Our observed cumulative number of halos follows the $\Lambda$CDM prediction 
(dashed line) for halo masses $> 5 \times 10^9$ M$_{\odot}$. This is in contrast with the observed abundance in the Milky Way. 
Uncertainties  due to uncertainties in the recovered shape of the matter power spectrum are not visible on this plot.
Therefore, even in this example study on a relatively small data set, nearly up to Sagittarius mass scales there seems to be no lack of observed power  in the CFHTLenS volume. In order to explore much smaller masses ($< 10^9 M_{\odot}$) we will need to extend our analysis to larger $k$ i.e., $k \sim 100$, however 
a straightforward  extension of our approach to these small scales is computationally prohibitive (computation time scales like $k^3$);  we will investigate this in future work.

\section{Discussion and Conclusions}
In this paper we have made the connection between inferences that are obtained from cosmic shear data on the total matter power spectrum on small-scales and the problem of the abundance of low mass galaxies. We point out that over large volumes, cosmic shear data can make a statistical statement on the small scale power and therefore on the issue of abundance of low mass 
galaxy fraction, and we use a recent result from current data as an example of the type of the analysis that can be used in this context. 

Using the reconstructed small scales shape of the matter power spectrum  obtained by \citet{KitchingI} from the CFHTLenS survey,  we have investigated 
if it is consistent with a suppression of small scale power, 
compared to the standard $\Lambda$CDM power spectrum,  sufficient to explain the ``missing satellite problem''.
The reconstructed power spectrum is used to infer the abundance of halos in the mass range $5 \times 10^9 < M_{\rm halo} < 10^{11}$. 
The $P(k)$  reconstruction in \citet{KitchingI} has substantial freedom at small scales, provided by the parameterisation of the baryonic feedback effects. 
The shape of the recovered power spectrum is allowed to deviate from the $\Lambda$CDM one by 3\% at the 95\% confidence level. This is  not sufficient to solve the missing satellites problem.

By using a  standard CDM model for the collapse of halos, our measurement of the power spectrum provides  estimates for the abundance of low mass 
galaxies assuming that gravity is the dominant force at play. To investigate the abundance of halos provided by this $P(k)$ we used the standard 
tool of the conditional mass function -- using the updated fitting formula by \cite{coyote}. We found that, on average, the predicted abundance 
of low mass halos is in agreement with the $\Lambda$CDM predictions down to masses of $\approx 5 \times 10^9$ M$_{\odot}$ in the CFHTLenS volume. In other words: a small-scale power suppression invoked 
to explain away the  missing low mass sub-halos in the Milky Way and the local group is not supported by current data.

Seventeen years after the  paper by  \citet{kamionkowskiLiddle00}, who suggested that a suppression of dark matter power on small scales could ease the $\Lambda$CDM dearth of dwarf galaxies  problem, we point out how the power on those scales can be measured  and a potential power suppression observationally constrained. In our approach,  by observing the evolved power spectrum, we can also in principle constrain models of warm dark matter and not only a change in the primordial power spectrum.

In our analysis we allow for, in principle, a massive, although not sharp or discontinuous, small scale suppression. We keep  the cosmology fixed to the \emph{Planck} values; this simply defines the power spectrum at large scales. At small scales, where it matters for 
the small halo abundance, the $P(k)$ reconstruction could have given us a massive suppression, but  it did not: 
the allowed suppression is a maximum of ${\cal O}$(1\%).

Most semi-analytic models of galaxy formation achieve a reconciliation between the observed and predicted abundance of low luminosity galaxies 
by drastically decreasing the baryon-to-dark matter fraction for faint galaxies. At present, there is no observational evidence from 
rotation curve modelling that low circular velocity disks are dark matter dominated (which would be the case if 
the baryon-to-dark matter fraction were very small). The alternative explanation is that our Local Group is a-typical 
and that its substructure abundance  does not correspond to the  mean of $\Lambda$CDM halos distribution, but as cosmological 
simulations of the Hubble volume show \citep{apostle} it is more a $3\sigma$  outlier. 

Whilst other recent cosmic shear results, e.g. \citet{2017arXiv170801538T,2017MNRAS.465.1454H}, 
do not include an explicit $k$-mode dependancy, which makes comparison 
more involved, they nevertheless find that their results are consistent at the cosmological parameter inference level with the \emph{Planck} cosmology 
\citep{2017arXiv170801530D,2017arXiv170700483E} without significant suppression of small-scale power. With the 
caveat that further study of these data is required in order to determine robust conclusions, 
this implies that these results would support the conclusions drawn from the smaller data set we used here. However some recent galaxy-galaxy lensing 
measurements suggest a suppression of small-scale power \citet{2017MNRAS.467.3024L}. Given that a decade of observational effort in weak lensing  surveys is coming to fruition, with major collaborations releasing the data (e.g., DES \citet{2017arXiv170801530D}, KiDS \citet{2015A&A...582A..62D}), it is timely to point out how these data could bear  on important open problems in cosmology and astrophysics, which may not have been among the original science drivers of the surveys.

In this short paper we have presented a first study on how to use the recovered matter power spectrum from cosmic shear data to constraint the 
abundance of small mass halos. Current and forthcoming surveys will provide a better control of systematic errors and cover larger volumes, 
thus allowing for a more thorough study of the small scale power spectrum of matter halos. 

\section*{Acknowledgments}
Funding for this work (RJ and LV) was partially provided by the Spanish MINECO under MDM-2014-0369 of ICCUB (Unidad de Excelencia ``Maria de Maeztu") and by MINECO grant AYA2014-58747-P AEI/FEDER UE. LV  acknowledges support of H2020 ERC 725327 BePreSysE project. TDK is supported by a Royal Society University Research Fellowship. We thank Alan Heavens for useful discussions.

\end{document}